# THE NLC SOFTWARE REQUIREMENTS METHODOLOGY

G.R. White, H. Shoaee, SLAC, Stanford, CA 94025, USA

Abstract

We describe the software requirements and development methodology developed for the NLC control system. Given the longevity of that project, and the likely geographical distribution of the collaborating engineers, the planned requirements management process is somewhat more formal than the norm in high energy physics projects. The short term goals of the requirements process are to accurately estimate costs, to decompose the problem, and to determine likely technologies. The long term goal is to enable a smooth transition from high level functional requirements to specific subsystem and component requirements for individual programmers, and to support distributed development. The methodology covers both ends of that life cycle. It covers both the analytical and documentary tools for software engineering, and project management support. This paper introduces the methodology, which is fully described in [1].

## 1 BACKGROUND AND OUTLINE

The estimated budget for NLC controls is about $0.5bnUS. The software development effort and support infrastructure is about $75m, and represents ~475 person years of effort. However the funding calendar requires that our staffing is not constant over the development period.

Additionally, the software controls effort is likely to be distributed and collaborative, involving EPICS developers and other participating institutions. Therefore a formal methodology has been developed which we hope will help to: 1) communicate future intent and alternatives, 2) define the "treaty points" between sub-systems, and the interfaces between sub-systems, and 3) help designers delineate how sub-systems relate to the network architecture and data infrastructure (such as data acquisition, control, database and archiving).

The NLC controls project also has a very long timeline and the objectives of our methodology change significantly over time. In the short term our objective is to effectively estimate the software and infrastructure requirements for the purposes of cost estimation and project scheduling. In the long term our goal shifts to producing requirements suitable for producing accurate and clear designs. This methodology is designed to help on both fronts. It separately outlines development and management methods, and is based on a combination of systematic structured methodologies [2][5], in particular the Rational Unified Process, but incorporates elements from systemic methods [4][7]. "Use Cases" (part of the Unified Modeling Language) are used to delineate user level requirements [6], but also significant emphasis is given to the handling of non-functional requirements.

## 2 METHODOLOGY FOR LARGE SYSTEM DEVELOPMENT

Clearly defined documentation deliverables are itemized and described explicitly by the methodology. Table 1 summarises those deliverables produced for each sub-system, application or component library; table 2 itemizes those for project management.

Table 1: Development Effort Deliverables

| Artifact | Description |
| --- | --- |
| Vision Document | Overall summary of purpose and desired features |
| Discussion Docs | Textual discussion of Concepts |
| Domain Model | Use Case Model: Diagrammatical and **Textual** description of Use Cases (that is, scenarios of use from a functional perspective). <br><br> Entity-Relationship diagrams. <br><br> Algorithms |
| Developers Guides | Continuously updated programmers documentation as systems take shape. |
| User Interface Model | User interface mock-up |
| Use Case Package Report: | Diagrams of relationships of packages and subsystems |
| Non-functional requirements Document | Textual description of system constraints and references |
| Conflict Matrix | How conflicting data acq or control requirement will be handled. |
| Taxonomy | Summary of key functional attributes |
| Requirements Matrices | RequisitePro db of requirements |
| Cost Benefit Analysis | Description of Cost, Effort and functionality alternatives and estimates |

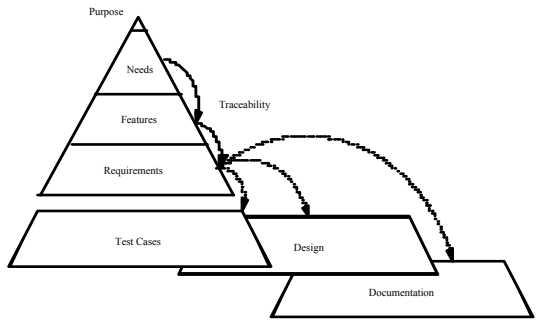

Figure 1: Hierarchy of Requirements

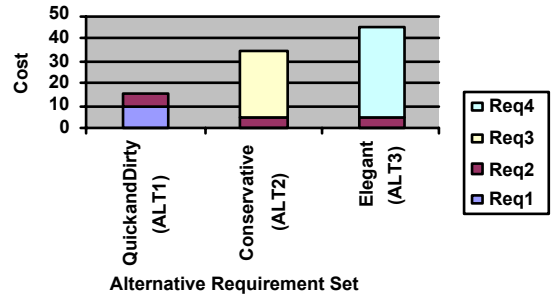

Figure 2: Comparison of Alternative RequirementSets

The methodology distinguishes between functional requirements (which describe the desired behavior of the system) and non-functional requirements (those concerned with the system's effective implementation). All requirements are viewed hierarchically; with very general requirements such as the purpose and goals at the top, followed by software features and then more specific, possibly testable items and "system requirements". As such, each requirement may be traced from its antecedents to its consequents (see Figure 1).

Each requirement is assigned an explicit description type, such as NEED, testable REQuirement, and so on (though absolutely correct classification is not necessary). For a very large distributed project, the role and specificity of *non-functional* requirements is particularly pronounced because these typically define the interfaces and shared resources of sub-systems. Table 3 lists some examples of these very specific non-functional requirements.

To each requirement are attached some "attributes" (different attributes depending on the requirement type) such as benefit, effort, cost, time, priority, risk of overrun etc, used for cost/benefit analysis.

One important attribute is obviously the level of "benefit" – captured as attribute values COULD, SHOULD and MUST. Additionally though we predict that in a large project there will be a large number of alternative requirements for alternative solutions proposed, each impacting the requirements of associated systems and changing their cost, time estimates, and level-of-benefit. Therefore the methodology proposes the documentation of "alternative requirement sets" to manage the treaty points between systems, their priority, cost alternative solutions, and to minimize friction and scope creep (see Figure 2).

To manage all this formal documentation we use a requirements management tool from Rational Software Corp. called RequisitePro, which is basically a database application. Each requirement in a document is captured using document parsing software. The resulting database of requirements can be queried, visualized, and analyzed with the usual drill-down and slice-and-dice paradigms [3].

Table 2: Global Deliverables

| Artifact | Description |
| --- | --- |
| CDR summary | Summary of requirements work breakdown |
| WBS | Work Breakdown Structure |
| Basic Design Principles | Statement of intent about the core architecture, infrastructure, and design of systems. |
| Sizing and Performance Specification | Diagrams and text showing nodes, data flow, and timing requirement |
| System Boundary Analysis | Summary of System Boundary and Treaty Points |
| Domain Analysis | Document and Diagrams detailing domain objects and their relationship |
| Enterprise db Schema | Description of control system entities and their relations |
| Glossary | A dictionary of terms. |
| References | Pointers to literature, both accl. Physics and Engineering and programming |

Table 3: Non-functional Requirement Types Traced

| Abbrev. | Description |
| --- | --- |
| XREF | Requirement of another system. |
| UC | A description of a use of the system, described from an actors point of view. |
| ALT | An alternative NEED, or FEAT, and REQ set which can satisfy the purpose. |
| PERF | Performance Requirement. |
| DBNAME | A database element, being a collection of attributes. This may describe a device, or some other aggregate. |
| TESTC | A test suite, aimed at testing a Feature. |

The global requirements analysis includes the construction of a "conflict matrix", which attempts to define the needs of each control subsystem (RF, feedback, beam monitor control etc) regarding their priority in contention situations with other systems. In the NLC this will be used define contention resolution rules.

The methodology includes a taxonomy for the key control system characteristics of each major subsystem, see Table 4.

Table 4: Controls Subsystem Taxonomy

| Property | Description |
|---|---|
| Initiation and Termination | How is the system started and terminated. |
| Duration | For how long is the system typically active and not idle |
| Frequency | How often is it initiated. |
| Read/write | Is the process mainly a reader or writer |
| Running State | What machine state is this system for? |
| Shareable input | Is the pulse related input data of this system shared by other systems. |

Data handling and network performance requirements are driven by data throughput and processing requirements. In order to accurately predict what those will be we constructed a "data sizing" taxonomy, including volume, frequency, latency, resilience, structure etc, which helps to identify the different data handling needs of each subsystem which is involved with the same data. For instance, the data handling need of a monitor display are very different to those of the monitor data acquisition system, or the monitor data archive system. The result is a logical network and storage architecture.

We propose not to have monolithic design, followed by development phases, as in the traditional "waterfall" approach. Rather we suggest an iterative, feature driven approach, with fixed 6-week periods in which we decide what features will be added or refined in each period. More analysis and design of each feature will be done in early iterations of the overall project, more programming and testing in later iterations. This is a formalized version of "extreme programming" which has recently become popular.

Over a long development period important design ideas and other incidentals, which cannot be pursued in the immediate term, are documented and tracked specifically as "forward references".

## 3. IMPLEMENTATION AND RESULTS

A formal systematic methodology was not used before in the SLAC controls community, though we had in the past applied the traditional approach very effectively, and design reviews and code standards were enforced. Although we have not started the NLC controls effort yet, we have adopted parts of this methodology for the existing B-factory, and our effort to migrate the existing controls system to new technologies (i.e., off VMS). In particular the formal capture of requirements into a db, the use of distinct vision, requirements, and discussion documents, forward references, iterative fixed period development, and continuously developed programmers documentation have all been implemented successfully and have increased productivity. We have also started data sizing the NLC, and created the taxonomy of subsystems. Some lessons have been learnt, and refined our plans for the methodology, which have been reflected in this paper. In particular, we had proposed a very structured workflow, with identified roles for each job function. That will probably be removed from the methodology user guide, which is available in [1].

This paper has introduced the major concepts of the methodology we developed. We would very much like to hear feedback on the methodology presented, and experiences of formal methodologies in use at other accelerator software departments.